\documentstyle[12pt,a4wide]{article}
\newcommand{\bra}[1]{\langle #1 |}       \newcommand{\ket}[1]{| #1 \rangle}
\newcommand{\ew}[1]{\langle #1 \rangle}
\newcommand{\ba}{\begin{array}}          \newcommand{\ea}{\end{array}}
\newcommand{\bea}{\begin{eqnarray}}      \newcommand{\eea}{\end{eqnarray}}
\newcommand{\beq}{\begin{equation}}       \newcommand{\eeq}{\end{equation}}

\def\roughly#1{\raise.3ex\hbox{$#1$\kern-.75em\lower1ex\hbox{$\sim$}}}
\newcounter{saveeqn}

\begin{document}
\thispagestyle{empty}
\hspace*{12cm}\parbox{3cm}{ITP-UH-19/93\\June 1993\\}

\noindent
{\large \bf The S=\boldmath$\frac{1}{2}$ Heisenberg antiferromagnet on the
triangular lattice:}\\
{\large \bf exact results and spin-wave theory for f\/inite cells}
\vspace*{1.5cm}

\noindent\hspace*{4cm}
R. Deutscher
\footnote{e-mail: deutsch@kastor.itp.uni-hannover.de;
\,telephone: 0511/762\,4836}
and H.U. Everts
\footnote{e-mail: everts@kastor.itp.uni-hannover.de;
\,telephone: 0511/762\,3886;\,FAX: 0511/762\,3023}

\noindent{\small \sl
Institut f\"ur Theoretische Physik, Universit\"at Hannover, Appelstr.~2,
D-3000 Hannover~1,}\\
{\small \sl  Federal Republic of Germany}\\

\noindent
We study the ground state properties of the S=$\frac{1}{2}$ Heisenberg
antiferromagnet (HAF) on the triangular lattice with nearest-neighbour ($J$)
and next-nearest neighbour ($\alpha J$) couplings. Classically, this system is
known to be ordered in a  $120^\circ$ N\'eel type state for values
$-\infty<\alpha\le 1/8$ of the ratio $\alpha$ of these couplings and in a
collinear state for $1/8<\alpha<1$. The order parameter ${\cal M}$ and the
helicity $\chi$ of the $120^\circ$ structure are obtained by numerical
diagonalisation of finite periodic systems of up to $N=30$ sites and by
applying the spin-wave (SW) approximation to the same finite systems. We
find a surprisingly good agreement between the exact and the SW results in
the entire region $-\infty<\alpha< 1/8$. It appears that the SW theory is still
valid for the simple triangular HAF ($\alpha=0$) although the sublattice
magnetisation ${\cal M}$
is substantially reduced from its classical value by quantum fluctuations.
Our numerical results for the order parameter ${\cal N}$ of the collinear
order support the previous conjecture of a first order transition
between the $120^\circ$ and the collinear order at $\alpha \simeq 1/8$.\\
PACS: 75.10J; 75.40M

\vspace*{5mm}
\noindent
{\sl to be published in $\qquad$ Z. Phys. B: Condensed Matter}
\vspace*{3mm}
\setcounter{page}{0}
\newcommand{\alpheqn}{\setcounter{saveeqn}{\value{equation}}%
\stepcounter{saveeqn}%
\setcounter{equation}{0}%
\renewcommand{\theequation}{\arabic{section}.
\mbox{\arabic{saveeqn}\alph{equation}}}}
\newcommand{\reseteqn}{\setcounter{equation}{\value{saveeqn}}%
\renewcommand{\theequation}{\arabic{section}.\arabic{equation}}}
\reseteqn
\section{Introduction}

Owing to the observation of quasi two dimensional antiferromagnetism in the
undoped phase of high $T_c$ superconductors \cite{HTC}, the theory of two
dimensional quantum antiferromagnetism has received much attention recently.
The continuing discussion focusses on the question of whether or to which
extent the classical picture of a N\'eel type ordered ground state remains
valid in the presence of quantum fluctuations which are known to be
particularly important in low dimensional systems. For the spin-$1/2$
isotropic Heisenberg antiferromagnet (HAF) with nearest neighbour interaction
on a square lattice this question has been settled: the result of numerous
quantum Monte Carlo (MC) simulations \cite{Barnes} leave no doubt about the
existence of N\'eel order in the ground state of the square lattice HAF.
Moreover, by refined MC techniques \cite{Wiese} it has been possible to
determine the staggered magnetisation ${\cal M}$ of this state  with high
accuracy. Quantum fluctuations are found to reduce ${\cal M}$ to about $60\%$
of its classical value, but surprisingly the relative difference
between the most accurate numerical value
${\cal M}=0.3074(4)$ and the value obtained in first order spin-wave (SW)
approximation ${\cal M}_{1.SW}=0.30340$ is only about $1\%$.
Long range order is opposed not only by quantum fluctuations but also by
frustration. For a spin-$1/2$ Heisenberg model, frustration is
defined in the sense introduced by
Toulouse \cite{Toul} for the z-z part of the coupling and, with respect
to the x-x and y-y parts of the interaction, by the impossibility of using the
Marshall Ansatz \cite{Marsh} for the ground state wave function \cite{Faze}.
The HAF
on a triangular lattice (HAFT) which will be studied in this work is the
prototypical example of a frustrated
antiferromagnet. In early work, Anderson and
Fazekas \cite{Andy1} suggested that the combined effect of quantum fluctuations
and of frustration would favour a disordered overall singlet state over an
ordered N\'eel type ground state. The state proposed by Anderson and
Fazekas consists of products of singlet pair states and is commonly referred
to as a resonating valence bond (RVB) state. It has received much attention
in the theory of high $T_c$ superconductivity. Numerous subsequent
investigations have led to contradictory views of the ground state of the
HAFT. The spin-wave analysis \cite{Joli1,Miya1}
predicts that quantum fluctuations are
insufficient to suppress the classical N\'eel order. Certain variational
calculations support this conclusion \cite{Huse}, others contradict it.
In fact, an exotic RVB-type state which breaks time reversal invariance and
parity has also been obtained by a variational method \cite{Kalm}.
Several attempts have been made to determine the nature of the ground state
of the HAFT by extrapolating from results obtained by the exact numerical
diagonalisation of the Hamiltonian of small periodic cells \cite{Fuji,Runge}.
In all of these studies the results of the extrapolations have been
interpreted as being
indicatory of a disordered ground state or of a state that is close to the
critical point of losing long range order. The analysis of a high-order
perturbation expansion has lent support to this view \cite{Singh}.
However, by extending the
numerical studies of finite cells to include not only the ground state
energy but the entire spectrum of the HAFT and by an appropriate interpretation
of their results Bernu et. al. \cite{Bernu} have very recently given convincing
arguments in favour of a N\'eel ordered ground state of the HAFT. In view of
the experience with spin-$1/2$ square lattice HAF, it might thus be hoped that
the SW approximation also yields accurate results for the physical
quantities
of the HAFT. It is one of the purposes of this study to show  that the results
obtained by the exact numerical diagonalisation of small cells of the HAFT
do indeed point towards the validity of the SW approximation.

With regard to this purpose it will prove useful to consider the HAFT with
nearest ($nn$) and next nearest neightbour ($nnn$) interaction,
\bea
H&=&\sum_{nn}^N {\bf S}_i {\bf S}_j + \alpha \sum_{nnn}^N {\bf S}_i {\bf S}_j
\label{Hamiltonian}
\eea
where ${\bf S}_i$ denotes a spin-$1/2$ operator at the lattice site
${\bf R}_i$.
(We set the $nn$ coupling equal to unity without loss of generality).
As is seen in Fig.~1,
the $nnn$ interaction couples spin pairs within each of the
three sublattices. Therefore, for sufficiently strong
ferromagnetic $nnn$ coupling,
$-\alpha\gg1$, the spins of each sublattice will order ferromagnetically,
forming a macroscopic spin. In this situation, a SW expansion will certainly be
applicable. Thus, the parameter region $-\alpha\gg1$ of the model
(\ref{Hamiltonian}) provides a firm basis for a comparsion of the SW
approximation and of the exact numerical result.

The model (\ref{Hamiltonian}) is also of considerable interest in its own
right. In the classical approximation one finds that its ground state changes
discontinuously from the $120^\circ$ three sublattice state for
$\alpha<\alpha_{cl}=1/8$ to a four sublattice state for $\alpha_{cl}<\alpha<1$
\cite{Joli3}, Fig.~2. The four-sublattice state is degenerate with respect to
the angle $\theta$, i.e. its energy is $E=-2(1+\alpha)$ independent of
$\theta$.
As has been shown in Ref. \cite{Joli3} quantum fluctuations select
the state with
$\theta=0$ from the continuum of degenerate four sublattice states.
We shall refer to this state, which is a two sublattice structure, as the
collinear state.
At $\alpha=1$ the ground state structure changes again to become
incommensurate for  $\alpha>1$. In the limit  $\alpha\to \infty$, the three
sublattices decouple so that a $120^\circ$ structure exists on each of them.
The manifold of classical ground states of (\ref{Hamiltonian})
that evolves as $\alpha$ increases from $-\infty$ to  $\infty$ is concisely
described by
\bea
{\bf S}_i = \hat{{\bf e}}_1 \,\cos {\bf Q}{\bf R}_i+\hat{{\bf e}}_2\,
\sin {\bf Q}{\bf R}_i ,
\label{KlassKonfig}
\eea
where $\hat{e}_1$ and $\hat{e}_2$ are an arbitary pair of orthogonal unit
vectors and where ${\bf Q}={\bf Q}(\alpha)$ traces out the path in the
Brillouin zone shown in Fig.~3.

The question of how the transition between the $120^\circ$ and the collinear
state of the HAFT with $nn$ and $nnn$ couplings is affected by
quantum fluctuations has attracted considerable interest
\cite{Joli3,Baska,Deut,Joli2}. If both types of long range order survive at the
quantum level, the transition between them may be of first order, i.e.
the order parameters of both states may remain finite at the transition point
so that the symmetry breaking pattern changes discontinuously.
Alternatively, one or both types of long range order could be suppressed in the
vicinity of the classical transition point, and an intermediate ordered or
disordered state could appear at the quantum level.
Disregarding the first possibility,
Baskaran conjectured that a time reversal and parity breaking
chiral spin liquid state should evolve as the ground state of the model
(\ref{Hamiltonian}) when $\alpha$ exceeds a critical value \cite{Baska}.
In a previous study \cite{Deut}, we could not confirm this conjecture. More
recently, Chubukov and Jolic{\oe}ur \cite{Joli2} investigated the collinear
structure by a self-consistent SW approximation. These authors find that the
stability region of the collinear structure extends to values of $\alpha<
\alpha_{cl}$ and argue that the transition between the $120^\circ$ and the
collinear structure should be of first order.
It is the second purpose of this paper to provide and discuss further
results concerning this transition.

In section 2 we work out the second
order SW expansion for physical properties of the HAFT pertaining to the
$120^\circ$ structure, and we show how finite size results are obtained within
the SW approximation. Section 3 contains our numerical results for
physical properties that characterise the ordered states of our model
(\ref{Hamiltonian}): the sublattice magnetisations, the helicity and the
chirality. The emphasis is on a direct comparison of the exact numerical
results with the predictions of the SW-theory for finite systems
and on the discussion of the
behaviour of the order parameters in the transition region between the
$120^\circ$ structure and the collinear structure. In section 4 we summarise
our conclusions. Technical details are deferred to three appendices.
\setcounter{equation}{0}
\section{Spin-wave approximation}

To order $1/S$, the spin-wave approximation for the model (\ref{Hamiltonian})
has been discussed by Jolic{\oe}ur et al \cite{Joli1} and a calculation of the
$1/S^2$ corrections to the spin-wave spectrum of the collinear state has been
performed by Chubukov and Jolic{\oe}ur \cite{Joli2}. Furthermore, for
$\alpha=0$ the $1/S^2$ correction to the sublattice magnetisation has been
obtained by Miyake \cite{Miya1}. Below we present the results of an extension
of Miyake's calculation to the range $-\infty\ge\alpha\ge 1/8$ where the
$120^\circ$ state is supposedly stable. With the objective of comparing the
SW approximation with our numerical results we also include the first and
second order results for the helicity of the $120^\circ$ structure, and we
discuss the finite size corrections to the sublattice magnetisation and the
helicity.

The starting point of the SW expansion is the Hamiltonian
\bea
H&=&\sum_{nn} H_{i j}+\alpha\,\sum_{nnn} H_{i j} - B \sum_i \tilde{S}_i^z,
\nonumber\\
H_{i j}&=&\cos \theta_{i j}\,(\tilde{S}_i^x \tilde{S}_j^x+
\tilde{S}_i^z \tilde{S}_j^z)+
\sin \theta_{i j}\,(\tilde{S}_i^z \tilde{S}_j^x-
\tilde{S}_i^x \tilde{S}_j^z) + \tilde{S}_i^y \tilde{S}_j^y
\label{SWHamiltonian}\\
\mbox{with}\quad\theta_{i j}&=&\theta_j - \theta_j,\nonumber
\eea
which is obtained from (\ref{Hamiltonian}) by rotating the local z-axis
to the direction of the classical sublattice magnetisation:
\bea
S_i^x=\cos \theta_i \,\tilde{S}_i^x + \sin \theta_i \,\tilde{S}_i^z,\quad
S_i^y=\tilde{S}_i^y,\quad
S_i^z=\cos \theta_i \,\tilde{S}_i^z - \sin \theta_i \,\tilde{S}_i^x.
\label{Dreh}
\eea
For the $120^\circ$ state $\theta_i=0,2\pi/3,-2\pi/3$, when $i$ is a site of
the sublattice ${\sf A,B}$ or ${\sf C}$ while $\theta_i=0,\pi$ on
the sites of the
two sublattice ${\sf A}$ or ${\sf B}$ of the collinear state.
$B$ represents a staggered magnetic field.
{}From (\ref{SWHamiltonian}) the SW expansion is obtained by replacing
the spin-operators $\tilde{S}_i^{\alpha}$ by Holstein-Primakoff boson
operators and by expanding with respect to $1/S$. In the case of the
$120^\circ$ state the result is
\bea
H &=& S^2\,(H_0 + H_1 + H_{3/2} + H_2) ,
\eea
with
\alpheqn
\bea
H_0 &=& -\frac{3}{2} (1+2\alpha) N -\frac{B}{S} N ,\\
H_1 &=& -\frac{1}{S} \sum_{nn} \left\{ -\frac{1}{4}(a_i^+a_j + h.c.)
+\frac{3}{4}(a_i^+ a_j^+ + h.c.)- \frac{1}{2}(n_i + n_j) \right\}\nonumber\\
&& + \frac{B}{S} \sum_i n_i +
\frac{\alpha}{S} \sum_{nnn} \left\{(a_i^+a_j + h.c.)-(n_i+n_j)
\right\},\\
H_{3/2} &=& -\sqrt{\frac{2}{S^3}} \sum_{nn} \sin \theta_{i j}
( n_i a_j + h.c.),\\
H_2  &=& -\frac{1}{S^2} \sum_{nn} \left\{ \frac{1}{2}(n_i n_j)+
 \frac{1}{8}(n_i a_i a_j^+ + h.c.)-\frac{3}{8}(n_i a_i a_j + h.c.)
\right\}\nonumber\label{SWH2}
\\
&&+\frac{\alpha}{S^2} \sum_{nnn} \left\{n_i n_j - (n_i a_i a_j^+ + h.c.)
\right\}.
\eea
\reseteqn
Diagonalisation of $H_1$ yields the SW frequency
(we adopt the notation of Miyake \cite{Miya1} wherever possible):
\bea
\omega({\bf k},B)&=& 3S\,\nu({\bf k},B)\nonumber\\
 &=& 3S\sqrt{
\left(1+\frac{B}{3S}-2\alpha({\bf k})-\gamma({\bf k})\right)
\left(1+\frac{B}{3S}-2\alpha({\bf k})+2\gamma({\bf k})\right)},
\label{SWfrequ}
\eea
\alpheqn
where
\bea
\alpha({\bf k})&=&\alpha-\frac{\alpha}{3}\left(\cos(\sqrt{3}k_y) +
\cos(\frac{3 k_x}{2}+\frac{\sqrt{3} k_y}{2})+
\cos(\frac{3 k_x}{2}-\frac{\sqrt{3} k_y}{2})\right)\label{Alphak},\\
\gamma({\bf k})&=&\frac{1}{3}\left(\cos(k_x) +
\cos(\frac{k_x}{2}+\frac{\sqrt{3} k_y}{2})+
\cos(\frac{k_x}{2}-\frac{\sqrt{3} k_y}{2})\right)\label{Betak}.
\eea
\reseteqn
For $\alpha<1/8$, $\omega({\bf k},0)\equiv\omega({\bf k})$
vanishes linearly in the centre of the
Brillouin zone (BZ) and at opposite corners
${\bf k}=\pm{\bf Q}_1=\pm(4\pi/3,0)$.
The zero modes are the three Goldstone modes of the
symmetry broken state \cite{Apel1}. The two modes at $\pm{\bf Q}_1$, which are
degenerate, correspond to an infinitesimal tilt of adjacent plaquettes
of three spins around two orthogonal axes in the plane defined by the
ideally ordered classical state. The third mode at ${\bf k}=0$ whose
SW velocity is larger than that of the previous ones by a factor
of $\sqrt{2}$ \,corresponds to an infinitesimal twist of adjacent plaquettes
around an axis that is perpendicular to the plane defined by the $120^\circ$
state.

As $\alpha$ approaches the value $1/8$ from below, the SW
frequency softens at ${\bf Q}^{(1)}_2,{\bf Q}^{(2)}_2$ and ${\bf Q}^{(3)}_2$,
Fig.~3, yielding new linear excitation
branches at these points. Within this first order approximation, these new
modes indicate the onset of collinear order at $\alpha=1/8$.
They are, however, unphysical since they are not Goldstone modes of the
$120^\circ$ state.

In first order in $1/S$ the sublattice magnetisation is given by
the expression \cite{Miya1}
\bea
{\cal M}_{1.SW} = S\,\left\{1+\frac{1}{4S}\,
\frac{1}{N} \sum_{{\bf k}}(2-A_1({\bf k})-A_2({\bf k}))\right\},
\label{Mag1stSW}
\eea
where
\bea
A_1({\bf k})=\sqrt{\frac{1-2\alpha({\bf k})+2\gamma({\bf k})}
{1-2\alpha({\bf k})-\gamma({\bf k})}},\quad A_2({\bf k})=A_1({\bf k})^{-1}.
\nonumber\eea
To obtain the second order correction ${\cal M}_{2.SW}$ we follow the method
of Miyake and calculate the $1/S^2$ correction to the ground state energy
\bea
E_0^{(2)}(\alpha,B)=S^2\ew{H_2}_0 +S^2 \sum_{\{n_{{\bf k}}\}}
\frac{|\bra{ \{n_{{\bf k}}\} }H_{3/2}  \ket{ \{0\} }|^2}
{E(\{0\})-E(\{n_{{\bf k}}\})},
\label{2.OrderSWEnergy}
\eea
where $\{n_{{\bf k}}\}$ is an eigenstate of $H_1$ and $E(\{n_{{\bf k}}\})$
is the corresponding energy
\bea
E(\{n_{{\bf k}}\})=\sum_{{\bf k}} (n_{{\bf k}}+\frac{1}{2})\, \omega({\bf k}).
\label{Energy2Ord}
\eea
The explicit expressions are given in the App.~A. As in the case $\alpha=0$
considered by Miyake, both contributions to $E_0^{(2)}(\alpha,B)$ contain
singular parts proportional to $\sqrt{B}$ which, however, can be shown to
cancel in the sum (\ref{2.OrderSWEnergy}). This feature has been utilised to
control the precision of the numerical computation of the integrals
contributing to $E_0^{(2)}(\alpha,B)$ and to extract the second order
correction ${\cal M}_{2.SW}$ to the magnetisation from  $E_0^{(2)}(\alpha,B)$
in the small $B$ limit,
\bea
{\cal M}_{2.SW} = -\frac{1}{N}\,\lim_{B\to 0} \frac{E_0^{(2)}(\alpha,B)-
E_0^{(2)}(\alpha,0)}
{B}.
\label{2.OrdSWMag}
\eea
In Fig.~4a we show plots of ${\cal M}_{1.SW}$ and of the
sum ${\cal M}_{1.SW}+{\cal M}_{2.SW}$.
A plot of ${\cal M}_{2.SW}$ is included in Fig.~8.
Owing to the unphysical zeros which
develop in $\omega({\bf k})$ at ${\bf Q}^{(1)}_2,{\bf Q}^{(2)}_2,
{\bf Q}^{(3)}_2$ as $\alpha$ approaches $\alpha_{cl}$ the second order
contribution ${\cal M}_{2.SW}$ diverges at $\alpha=\alpha_{cl}$.
However, ${\cal M}_{2.SW}$  enhances the  sublattice magnetisation also for
$\alpha\roughly<0.05$ where the influence of the divergence is still small.
Thus, it appears that the stability region of the $120^\circ$ structure
extends beyond $\alpha_{cl}$ and hence overlaps with the stability region of
the collinear structure. This corroborates the argument of
Chubukov and Jolic{\oe}ur \cite{Joli2} according to which the transition
between the two structures should be of first order.

In the classical  $120^\circ$ state an Ising-like variable can be attached
to each triangular plaquette of the lattice \cite{Villa}. It takes the values
$\pm1$ depending on whether the spins on the three sites of the plaquette
turn clockwise or counterclockwise. For the quantum system the operator
associated with this variable - we shall call it helicity -
has been defined by
\bea
\hat{\mbox{\boldmath$\chi$}}_{\bigtriangleup} =
\frac{2}{\sqrt{3}}\,({\bf S}_i \times {\bf S}_j +
{\bf S}_j \times {\bf S}_k +{\bf S}_k \times {\bf S}_i ),
\label{DefHel}
\eea
where $i,j$ and $k$ denote the sites of an elementary triangular plaquette
in a clockwise sense. The normalisation has been chosen such that
each component of
$\hat{\mbox{\boldmath$\chi$}}_{\bigtriangleup}$
has eigenvalues $+1,0$ and $-1$.
Long range helicity order is
measured by the expectation value \cite{Fuji2}
\bea
\chi^y
=\frac{1}{N}\,\sum_{\bigtriangleup}\ew{\hat{\chi}^y_{\bigtriangleup}},
\label{HelSum}
\eea
where the sum extends over all upward pointing triangles of the lattice.
Long range helicity order can exist in the absence of long range
$120^\circ$ order.
In that case it is the signature of a spin nematic \cite{Andr}.
It must necessarily
exist for a state  that possesses long range $120^\circ$ order.
In the local reference frame (\ref{Dreh}) of the  $120^\circ$ state,
${\bf S}_i \times {\bf S}_j$
takes the form
\bea
({\bf S}_i \times {\bf S}_j) \cdot \hat{\bf e}_y=
\sin\theta_{ij}\,(\tilde{S}_i^x \tilde{S}_j^x+\tilde{S}_i^z \tilde{S}_j^z)+
\cos\theta_{ij}\,(\tilde{S}_i^x \tilde{S}_j^z-\tilde{S}_i^z \tilde{S}_j^x).
\eea
{}From this expression one immediately obtains the order parameter $\chi^y$
in first order SW approximation
\bea
\chi^y_{1.SW}(\alpha)&=&3 S^2
\left\{1+\frac{1}{2S}\,\frac{1}{N} \sum_{{\bf k}}
(2-A_1({\bf k})-A_2({\bf k})-B_1({\bf k}))\right\}\nonumber\\
&=& 3 S^2 \left\{1+\frac{2}{S}({\cal M}_{1.SW}-S)+
\frac{1}{2S^3}\ew{\tilde{S}_i^x \tilde{S}_j^x}_{1.SW}\right\},
\label{Hel1stSW}
\eea
where
\bea
B_1({\bf k}) = \gamma({\bf k}) A_1({\bf k}).
\eea
A plot of $\chi_{1.SW}(\alpha)$ is contained in Fig.~4b.
As is described in App.~A, the second order correction  $\chi_{2.SW}$
can be obtained by the same technique as has been applied in calculating
${\cal M}_{2.SW}$.\, A plot of $\chi_{2.SW}(\alpha)$ is included in Fig.~10.
For $\alpha\roughly<-1$, the quantum fluctuations are seen to
enhance $\chi^y$ slightly over the classical value.
Quite generally, the quantum fluctuations are seen to be less effective
in reducing the helicity $\chi^y$ than in reducing the magnetisation
${\cal M}$. This is not unexpected, since the long wavelength modes  that
twist the spin configuration around the axis perpendicular
to the plane of the $120^\circ$ structure are ineffective in reducing
$\chi^y$.

As we have explained above we need to evaluate the SW expressions
(\ref{Mag1stSW}) and (\ref{Hel1stSW}) for finite periodic systems. This
requires modifications, since in (\ref{Mag1stSW}) and (\ref{Hel1stSW}) the
members of the ${\bf k}$ sums that correspond to the zero modes ${\bf k}=0$
and  ${\bf k}=\pm {\bf Q}_1$ are infinite.
In his seminal work on quantum antiferromagnetism, Anderson \cite{Andy2}
has pointed out that these infinities are unphysical, yet unavoidable
consequences of the basic assumptions underlying the SW theory. In previous
work on finite size effects \cite{Neub} it has been shown that the leading
size correction to ${\cal M}_{1.SW}(0,N)$ is obtained by omitting the
infinite contributions $A_1({\bf k}=0)$ and $A_2({\bf k}=\pm {\bf Q}_1)$
from the sum in (\ref{Mag1stSW}). As we show in App.~B this omission yields
an unphysical result in the limit of strong ferromagnetic $nnn$ coupling,
$-\alpha /N \gg 1$,
\bea
\lim_{-\alpha /N\to \infty} {\cal M}_{1.SW}(\alpha,N)=S\left\{1+\frac{1}{2S}
\frac{3}{N}\right\}>S.
\eea
At the same time we find that the expression
\bea
{\cal M}_{1.SW}(\alpha,N)&=& S
\left\{1+\frac{1}{4S}\,\frac{1}{N}
\sum_{{\bf K}\epsilon BZ \atop {\bf k}\neq 0,\pm {\bf Q}_1}
(2-A_1({\bf k})-A_2({\bf k}))\right\},
\label{MagFinSW}
\eea
in which not only the infinite parts of the zero mode contributions but also
the finite parts have been neglected, yields the correct result
\bea
\lim_{-\alpha /N\to \infty} {\cal M}_{1.SW}(\alpha,N)=S.
\eea
Furthermore, for finite $\alpha$ the leading $1/\sqrt{N}$ size correction
remains unaffected by the neglect of the finite parts of the zero mode
contributions.
Thus, (\ref{MagFinSW}) qualifies as a finite size approximant for
${\cal M}_{1.SW}$ in the entire region $-\infty\le\alpha\le 1/8$ where
the $120^\circ$ structure is classically stable.
Of course, for a specified finite periodic system of size $N$ the sum
(\ref{MagFinSW}) has to be extended over that discrete set of inequivalent
${\bf k}$-vectors which represents the system in the ${\bf k}$ space.
Plots of ${\cal M}_{1.SW}(0,N)$
obtained in this way and of the asymptotics
\bea
{\cal M}_{1.SW}(0,N)={\cal M}_{1.SW}(0,\infty)+\frac{1.215}{\sqrt{N}}
\label{MagAsym}
\eea
are shown in Fig.~5.
Obviously, for $N\roughly<10^2$ the subleading $N^{-1}$ corrections preclude
the use of the asymptotics (\ref{MagAsym}) in an extrapolation to the bulk
limit. Fig.~5 also contains a plot of
\bea
{\cal M}_{1.SW}(0,N)={\cal M}_{1.SW}(0,\infty)+\frac{1.215}{\sqrt{N}}
-\frac{3}{2N},
\label{Mag2Asym}
\eea
in which the last term is the subleading correction that obtains in the limit
$-\alpha/N\gg 1$, App.~B.
Evidently, (\ref{Mag2Asym}) provides a good approximation for
${\cal M}_{1.SW}(0,N)$. The remaining fluctuations in the data points reflect
the dependence of ${\cal M}_{1.SW}(0,N)$ on the shape of the finite cells,
i.e. on the specific set of ${\bf k}$ points to be used in (\ref{MagFinSW}).

By analogy with (\ref{MagFinSW}) we define the finite size approximants for the
helicity by
\bea
\chi^y_{1.SW}(\alpha,N)&=&3 S^2
\left\{1+\frac{1}{2S}\,\frac{1}{N}
\sum_{{\bf K}\epsilon BZ \atop {\bf k}\neq 0,\pm {\bf Q}_1}
(2-A_1({\bf k})-A_2({\bf k})+B_1({\bf k}))\right\}.
\label{HelFinSW}
\eea
In principle, finite size values for the second order quantities
${\cal M}_{2.SW}$ and $\chi_{2.SW}$ can also be obtained by omitting the
contributions of the zero modes. However, in practice the results for
$E_0^{(2)}(\alpha,B,N)$ turn out to depend irregularly on $\sqrt{B}$ so that
the extrapolation (\ref{2.OrdSWMag}) is not feasible.
\setcounter{equation}{0}
\section{Numerical results}

Presently, exact results for the HAFT are available for system sizes up to
$N=36$ \cite{Runge,Bernu}. Thus, in view of the results of the last section
there is little use in attempting to extrapolate from such system sizes to
the bulk properties of the HAFT. Clearly, however, one expects to find
agreement between the exact results and finite size SW results of the
previous section, at least in the region $-\alpha\gg 1$ where the
SW approximation has a firm basis.

In the following subsections we shall discuss the order parameters ${\cal M}$
and ${\cal N}$ of the $120^\circ$ structure and of the collinear structure
(subsections 3.1a and 3.1b).
The helicity $\chi$ and the chirality $P$ will be studied in subsections
3.2 and 3.3.  The numerical calculations include cells of sizes
$N=12, 18, 21, 24, 27$ and $30$ with periodic boundary conditions,
Figs.~6a-f.

The translationally invariant $N=21$ and $27$ cells
are fully symmetric under the $C_{6v}$
symmetry group of the Hamiltonian but they do
not accommodate the $2\times 1$ cell of the order parameter of the collinear
state. Since we are interested in obtaining both the order parameter of
the $120^\circ$ state, whose unit cell is of size $\sqrt{3} \times \sqrt{3}$,
and that of the collinear phase we include the $N=18,24$ and $30$ cells,
which accommodate both unit cells but are not symmetric under all $C_{6v}$
operations. We diagonalise the Hamiltonian (\ref{Hamiltonian}) for these
finite systems by using the Lanczos algorithm.
For $\alpha< 1/8$, the ground state of cells with an even number of sites are
translationally invariant singlets, $S_{tot}=0$, ${\bf Q}={\bf Q}_0=0$,
while the ground state of cells with an odd number sites are doublets,
$S_{tot}=1/2$, with wavevector ${\bf Q}_1=(4\pi/3,0)$.
At some value $\alpha>1/8$ the ground state of the $N=18$ and the $N=30$
cells cross over to the wavevector ${\bf Q}_2$,
which is the Bragg vector of the collinear state. As has been discussed in
\cite{Deut} for the case $N=21$, when $\alpha$ exceeds a
critical value $\alpha_{qu}(N)>1/8$
($\alpha_{qu}(21)=0.253,\alpha_{qu}(27)=0.130$) the ground
state of cells with odd $N$ cross over to a wavevector ${\bf Q}\ne
{\bf Q}_2$, since the periodicity of the collinear state is incompatible
with the periodicity of these cells.
We notice that the same type of behaviour already occurs at the classical
level. For the $N=21$ and the $N=27$ cell the classical ground state
(\ref{KlassKonfig}) jumps from ${\bf Q}={\bf Q}_1$ to a
${\bf Q} \neq {\bf Q}_2$ at $\alpha_{cl}(21)=0.173$ and
$\alpha_{cl}(27)=0.129$.

\subsection{Sublattice magnetisation}

{\bf 3.1a \quad The \boldmath$120^\circ$ order}

As a measure for the sublattice magnetisation ${\cal M}$ of the $120^\circ$
structure we calculate the structure function
\bea
S_N ({\bf Q}_{1})=
\sum_{i,j}^N e^{i {\bf Q}_1 ({\bf r}_i -{\bf r}_j)}\ew{ {\bf S}_i {\bf S}_j}_0
\quad\mbox{with}\quad{\bf Q}_1=(4\pi/3,0)
\label{StrukFkt}
\eea
from our numerically determined ground states.
To find the connection between $S_N ({\bf Q}_{1})$ and the finite size
approximants ${\cal M}(\alpha,N)$ for the sublattice magnetisation
we consider the
limit of large ferromagnetic $nnn$ coupling, $-\alpha\gg 1$. In this limit
the spins of each of the sublattices ${\sf A,B}$ and ${\sf C}$ will
align ferromagnetically forming macroscopic spins of magnitude
$S_{\sf A}=S_{\sf B}=S_{\sf C}=N/6$. These macroscopic spins may fluctuate
against each other so that the total spin
$S_{tot}=|{\bf S}_{\sf A}+{\bf S}_{\sf B}+{\bf S}_{\sf C}|$ agrees with the
values found in the
numerical calculation, i.e. $S_{tot}(N)=0$ for even $N$ and $S_{tot}(N)=1/2$
for odd $N$. A simple calculation which is detailed in App.~C yields
\bea
\lim_{\alpha\to -\infty} \frac{S_N ({\bf Q}_{1})}{N^2}=\left\{
\begin{array}{l@{\quad,\quad}l}
\frac{1}{8} + \frac{3}{4N} & N\, \mbox{even},\\
\frac{1}{8} + \frac{3}{4N} -  \frac{3}{8N^2}& N\, \mbox{odd}.
\end{array} \right.
\label{StrukSize}
\eea
Since
$$
\lim_{\alpha\to -\infty} {\cal M}(\alpha,N)
=\frac{3}{N}\,S_{\sf A}=\frac{3}{N}\,S_{\sf B}=\frac{3}{N}\,S_{\sf C}
=\frac{1}{2}
$$
it follows from (\ref{StrukSize}) that in the limit $\alpha\to -\infty$
\bea
{\cal M}^2(\alpha,N) = \left\{
\begin{array}{l@{\quad,\quad}l}
2\,\left(\frac{S_N ({\bf Q}_{1})}{N^2}- \frac{3}{4N}\right) & N\,
\mbox{even},\\
2\,\left(\frac{S_N ({\bf Q}_{1})}{N^2}- \frac{3}{4N} + \frac{3}{8N^2}\right)&
N\, \mbox{odd}.
\end{array} \right.
\label{MagSize}
\eea
Here, the factor of 2 reflects the fact that $S_N ({\bf Q}_{1})$ is the
structure factor of only one of the two $120^\circ$ structures with wave
vectors $\pm{\bf Q}_{1}$ which are contained in the ground state with equal
weight.
The above considerations suggest that (\ref{MagSize}) still holds
approximately for $-\alpha\roughly> 1$. The same argument also suggests that
for $-\alpha\roughly> 1$ the numerically exact results ${\cal M}(\alpha,N)$
should be well approximated by the finite size SW results
${\cal M}_{SW}(\alpha,N)$.
Plots of ${\cal M}_{1.SW}^2(\alpha,N)$, i.e. of the first order
SW result (\ref{MagFinSW}), and of ${\cal M}^2(\alpha,N)$ as functions of
$\alpha$ are shown in
Figs.~7a-f. The difference
\bea
\Delta {\cal M}(\alpha,N) ={\cal M}(\alpha,N)-{\cal M}_{1.SW}(\alpha,N)
\eea
is displayed in Fig.~8 together with the second order SW correction
${\cal M}_{2.SW}$. Obviously, the asymmetric $N=18$ cell yields
exceptionally large values for $\Delta {\cal M}$. Apart from this exception,
the first order SW approximation is seen to approximate the exact
results remarkably well even for small values of $|\alpha|$ where
quantum fluctuations evidently lead to a considerable reduction of ${\cal M}$
from its classical value.
Notably, for negative values of $\alpha$ up to $\alpha\simeq -0.1$,
the difference $\Delta{\cal M}$ depends only weakly on the system size $N$.
In this region, the second order correction ${\cal M}_{2.SW}(\alpha)$
overestimates the difference between the first order SW approximation and the
exact results. This is reminiscent of the SW expansion for the
$S=1/2$ square lattice
HAF in which the third order correction is found to reduce the discrepancy
between the exact results and the SW approximation by about $50\%$
\cite{Hamer}. For $\alpha \to \alpha_{cl}$, ${\cal M}_{2.SW}$ increases faster
than $\Delta{\cal M}$. It is tempting to attribute this discrepancy to the
unphysical divergence of ${\cal M}_{2.SW}(\alpha)$ at $\alpha_{cl}$. A
selfconsistent SW treatment \cite{Joli2} which eliminates this divergence is
certainly called for.

\vspace{1cm}
{\bf 3.1b \quad The collinear order}

As in the case of the $120^\circ$ order we use the
structure function $S_N ({\bf Q}_2)$ at the Bragg vector ${\bf Q}_2$
of the collinear structure as a measure for the order parameter $\cal N$.
If the collinear order of the ground states of the finite cells were
perfect the same steps that led to the relations
(\ref{MagSize}) would yield
\bea
{\cal N}^2(N) = f\left(\frac{S_N ({\bf Q}_2)}{N^2}-\frac{1}{N}\right).
\label{MagCollin}
\eea
The factor $f$ counts the number of inequivalent ${\bf Q}_2$ vectors in the
BZ. Hence, we get $f=3$ for the $C_{6v}$ symmetric $N=12$ cell, and
$f=1$ for the  $N=18,24,30$ cells which are not $C_{6v}$ symmetric.

As in the previous subsection we assume that the relation (\ref{MagCollin})
still holds approximately when the ground state is not perfectly ordered.
Figs.~7a,b,d and f show plots of ${\cal N}^2(\alpha,N)$ for
$\alpha\roughly>1/8$ obtained by employing the numerical values of
$S_N ({\bf Q}_2)$ in (\ref{MagCollin}). The maxima of ${\cal N}(\alpha,18)$
and ${\cal N}(\alpha,24)$ are seen to be only slightly reduced from the
saturation value ${\cal N}=1/2$. This lends support to our assumption about the
validity of (\ref{MagCollin}). The stronger reduction of
${\cal N}^2(\alpha,12)$ and the small discontinuity in ${\cal N}^2(\alpha,12)$
at $\alpha=0.225$ can be attributed to certain pecularities of the $N=12$
cell which will be discussed elsewhere \cite{Deut2}.
In the transition region $\alpha\simeq\alpha_{cl}$, the behaviour of
${\cal N}(\alpha,N)$ is seen to differ qualitatively between the
different system sizes. While ${\cal N}(\alpha,12)$ and ${\cal N}(\alpha,24)$
grow smoothly with $\alpha$, ${\cal N}(\alpha,18)$
increases discontinuously at $\alpha_{qu}(18)=0.23$.
This discontinuity is due to a crossover in the
ground state of the $N=18$ system whose translational symmetry changes from
${\bf Q}=0$ to ${\bf Q}={\bf Q}_2$ as $\alpha$ increases through
$\alpha=\alpha_{qu}(18)$.
In the classical picture (\ref{KlassKonfig}), the wavevector
${\bf Q}_2$ characterises
the periodicity of the collinear state so that for $\alpha>\alpha_{qu}(18)$
the ground
state of the $N=18$ system, $\ket{{\bf Q}_2}$, has the same periodicity as the
classical state. For the $N=30$ system we observe the same type of crossover
between ground states at $\alpha_{qu}(30)=0.2$.
Because of limitations in computer
memory we have not been able to calculate ${\cal N}(30)$ for
$\alpha>\alpha_{qu}(30)$
where we would have had to work with a translationally noninvariant ground
state
$\ket{{\bf Q}_2}$. One can be sure, however, that the  crossover
between ground states will also result in a discontinuity in ${\cal N}(30)$
at $\alpha=\alpha_{qu}(30)$.
The difference between the $N=12$ and $N=24$ systems, for
which the ground state remains at ${\bf Q}=0$ for all values of $\alpha$,
and the $N=18$ and $N=30$ systems is reminiscent of Marshall's rule
concerning the ground state of the square lattice HAF \cite{Marsh}:
the ground state is  translationally invariant, if the system size $N$
is a multiple of $4$; otherwise, when $N$ is even but not a multiple of $4$,
it has the  periodicity of the classical state.

As has been discussed in section 2 the SW theory suggests that the transition
between the $120^\circ$ structure and the collinear structure is of first
order. The discontinuities in the order parameter ${\cal N}$ of the
$N=18$ and $N=30$ systems directly confirm this prediction. For the system
sizes  $N=12$ and $N=24$, the plots of ${\cal N}$ and ${\cal M}$ are seen to
overlap in a finite interval around $\alpha=\alpha_{cl}$, Figs.~7a,d.
While the overlap region decreases when the system size is doubled
from  $N=12$ to $N=24$, the slopes of ${\cal N}$ and ${\cal M}$ are seen to
increase drastically at the transition points.
These features agree with the
behaviour of the order parameters of finite systems that exhibit
a first order transition in the bulk limit. In summary, our numerical
results support the view of a first order transition between the
 $120^\circ$ structure and the collinear structure of the HAFT.

\subsection{Helicity}

We wish to compare the SW results $\chi_{SW}^y(N)$ and the exact
values of the helicity. Numerically we calculate
\bea
\left\langle\left(\frac{1}{N}\sum_{\bigtriangleup}
\hat{\mbox{\boldmath$\chi$}}_{\bigtriangleup}\right)^2\right\rangle=3\,
\left\langle\left(\frac{1}{N}\sum_{\bigtriangleup}
\hat{\chi}^y_{\bigtriangleup}\right)^2\right\rangle,
\label{HelVec}
\eea
where $\hat{\mbox{\boldmath$\chi$}}_{\bigtriangleup}$ has been defined in
(\ref{DefHel}).
The identity (\ref{HelVec}) holds because the ground states are invariant under
rotations in spin space. To find the connection between $\chi_{SW}^y$
and
$\ew{(\sum_{\bigtriangleup}\hat{\mbox{\boldmath$\chi$}}_{\bigtriangleup})^2}$
we again consider the limit $\alpha\to -\infty$. Since the average
$\ew{(\sum_{\bigtriangleup}\hat{\mbox{\boldmath$\chi$}}_{\bigtriangleup})^2}$
contains four spin correlations an exact analytical evaluation is not
possible in the limit. By a mean field approximation we find
\bea
\lim_{\alpha\to -\infty}
\left\langle\left(\frac{1}{N}\sum_{\bigtriangleup}
\hat{\mbox{\boldmath$\chi$}}_{\bigtriangleup}\right)^2\right\rangle_{mf}=
9S^4+\frac{a_{mf}}{N}+ O(N^{-2}),
\label{HelMF}
\eea
with $a_{mf}=\frac{31}{4}$ for $S=\frac{1}{2}$.
If $\chi^y_{1.SW}(\alpha,N)$, (\ref{Hel1stSW}), is evaluated for finite $N$
one finds that the $S^{-1}$ corrections vanish in the limit
$\alpha\to -\infty$ so that from (\ref{Hel1stSW}) and (\ref{HelMF})
\bea
\lim_{\alpha\to -\infty} \left[\chi^y_{1.SW}(\alpha,N)\right]^2=
\lim_{\alpha\to -\infty}
\left\langle\left(\frac{1}{N}\sum_{\bigtriangleup}
\hat{\mbox{\boldmath$\chi$}}_{\bigtriangleup}\right)^2\right\rangle_{mf}
-\frac{a_{mf}}{N}+ O(N^{-2})=9S^4.
\eea
Instead of working with the approximate finite size correction $a_{mf}$ we
fit our exact results in the limit $\alpha\to -\infty$ by a second order
polynomial in $N^{-1}$,
\bea
\lim_{\alpha\to -\infty}
\left\langle\left(\frac{1}{N}\sum_{\bigtriangleup}
\hat{\mbox{\boldmath$\chi$}}_{\bigtriangleup}\right)^2\right\rangle_0=
9S^4+\frac{a}{N}+\frac{b}{N^2}.
\eea
The fit is done separately for even $N\;(N=18,24)$ and odd $N\;(N=21,27)$.
We find
\bea
a_{even}=7.63,\quad b_{even}=20.02;
\quad \quad a_{odd}=7.81,\quad b_{odd}=5.75\; .
\eea
Obviously, $a_{mf}=7.75$ is a good approximation for the leading finite size
correction. That the finite size corrections of the
helicity are large in comparison with the corresponding corrections
of the sublattice magnetisation (\ref{MagSize}) is not surprising
since each of the plaquet\-tes in
$\ew{(\sum_{\bigtriangleup}
\hat{\mbox{\boldmath$\chi$}}_{\bigtriangleup})^2}_0$ overlaps
with six neighbouring plaquettes.
In Figs.~9a-d we show plots of the numerical results
\bea
\chi^2(\alpha,N)=
\left\langle\left(\frac{1}{N}\sum_{\bigtriangleup}
\hat{\mbox{\boldmath$\chi$}}_{\bigtriangleup}\right)^2\right\rangle_0
-\frac{a}{N}-\frac{b}{N^2}
\eea
and of the SW approximations ${\chi}^2_{1.SW}(\alpha,N)$.
As is seen in Fig.~10, where we plot the
difference $\Delta {\chi}(N)={\chi}(N)-{\chi}_{1.SW}(N)$
between the first order SW results and the exact values for the
helicity, the agreement is not as good as in the case of the sublattice
magnetisation. The difference $\Delta {\chi}(N)$ deviates appreciably
from the second order SW approximation shown in Fig.~10 even for the
large $C_{6v}$ symmetric cells $N=21$ and $N=27$. Nevertheless, the enhancement
of the helicity over the classical value, Figs.~9a-d, in the region
$-\infty<\alpha\roughly<-0.2$,
which must be attributed to quantum fluctuations, is exhibited by both
${\chi}(N)$ and ${\chi}_{1.SW}(N)$ in a similar fashion.
Furthermore, the decrease of ${\chi}(N)$ and ${\chi}_{1.SW}(N)$
to zero as $\alpha$ approaches $\alpha_{cl}$ which is concomitant to the
decrease of the sublattice magnetisation ${\cal M}$ indicates clearly that the
helicity is just another measure for the $120^\circ$ long range spin order
of the HAFT and has no independent significance as in the case of a spin
nematic.

\subsection{Chirality}

As we have discussed in the introduction, a RVB state which breaks time
reversal and parity invariance has been proposed as the ground state
of the HAFT \cite{Deut}. Chiral symmetry breaking, i.e. a nonvanishing
expectation value of the operator ${\bf S}_i({\bf S}_j \times {\bf S}_k)$
defined on an elementary triangular plaquette with sites $i,j,k$, is known
to be the signature of such a state \cite{Wen}. According to Baskaran
\cite{Baska} chiral symmetry breaking should only occur when the $nnn$
coupling $\alpha$ exceeds a critical value.

In Fig.~11 we plot the expectation value
\bea
P_\pm = \frac{1}{N^2}\ew{C_\pm^+ C_\pm},
\eea
where $C_\pm$ are the uniform and staggered chiralities
\alpheqn
\bea
C_+ = \frac{2}{\sqrt{3}}\,
\sum_{\hbox{$\bigtriangleup$}\kern-.5em\raise.5ex\hbox{$\bigtriangledown$}}
\,\Big({\bf S}_i({\bf S}_j \times {\bf S}_k)
+
{\bf S}_j({\bf S}_l \times {\bf S}_k)\Big),
\label{Chi}
\\
C_- = \sqrt{2}\,
\sum_{\hbox{$\bigtriangleup$}\kern-.5em\raise.5ex\hbox{$\bigtriangledown$}}
\,\Big({\bf S}_i({\bf S}_j \times {\bf S}_k)
-
{\bf S}_j({\bf S}_l \times {\bf S}_k)\Big)
{}.
\eea
\reseteqn
Here, $i,j,k$ and $l$ denote the sites of a four-spin plaquette
as depicted in Fig.~12,
and the sums extend over the set of such plaquettes which covers the lattice.
The normalisation in (\ref{Chi},b) is chosen such that the maximum
modulus of the eigenvalues is unity in both cases. Contrary to Baskaran's
conjecture and in agreement with the picture of the ground state as it has
emerged so far no significant increase with $\alpha$ of either $P_+$ or
$P_-$ is observed.

\section{Summary and Conclusion}

In this work two important tools for investigating the ground state
structure of quantum antiferromagnets, spin-wave theory
and numerical diagonalisation of small periodic systems, have been applied
to the HAFT with $nn$ and $nnn$ interactions. Our spin wave calculations
extend previous work \cite{Miya1} on the  $120^\circ$ structure of the
model in various ways.
The extension of the SW-expansion of the
sublattice magnetisation ${\cal M}$ to order $S^{-2}$ appears to corroborate
the previous conjecture \cite{Joli2} that the
transition between the $120^\circ$ order and the collinear order of the HAFT is
of first order. With the aim of comparing the SW results with the exact
numerical results we develop finite size approximants for the former.
By the example of the first order SW approximation for the sublattice
magnetisation we demonstrate that the subleading $1/N$ size
corrections must not be neglected for system
sizes $N\roughly< 10^2$. Moreover, irregular corrections which reflect the
dependence of the finite size results on the shapes of the systems
are seen to be relevant for smaller sizes $N\roughly<36$. These results of the
SW approximation indicate that extrapolations from system size
$N\roughly<36$ which exclusively rely on the leading
$1/\sqrt{N}$ size dependence are unwarranted.
Therefore, in the discussion of our exact numerical results we do not attempt
to extrapolate to the bulk limit. Rather, we compare directly the finite size
data obtained by the SW approximation  with the corresponding exact data.

In our numerical investigation we calculate quantities which are suited to
characterise the $120^\circ$ order and the collinear order of the ground state
of the HAFT,
namely, the order parameter ${\cal M}$ and the helicity $\chi^y$
of the $120^\circ$
structure and the order parameter ${\cal N}$ of the collinear structure.
To check whether there is an indication for
unconventional order we also calculate the chirality $P$.
Incommensurate structures which evolve in the classical picture for
$\alpha>1$ are outside the scope of the present study since we work with
finite periodic systems.

For strong ferromagnetic $nnn$ coupling, $\alpha\ll-1$, one expects to find
aggreement between the exact results  ${\cal M}(\alpha,N)$ and
the first order SW approximation ${\cal M}_{1.SW}(\alpha,N)$ on physical
grounds. Suprisingly, however, we find that ${\cal M}_{1.SW}(\alpha,N)$
approximates ${\cal M}(\alpha,N)$ very well also for small values of
$|\alpha|$ where ${\cal M}(\alpha,N)$ is substantially reduced from its
classical value by quantum effects. This suggests that the SW
approximation remains valid even for small positive values of $\alpha$ where
the frustration already present in the triangular antiferromagnet is further
enhanced by the $nnn$ coupling. It appears that frustration is not as
detrimental to long range order in spin-$1/2$ quantum antiferromagnets as has
been thought \cite{Andy1,Runge,Sachd1}. The comparison of the
first order SW approximation for the helicity $\chi$ with the exact values
supports this conclusion:  $\chi$ is tied to the sublattice
magnetisation just as it should be for an ordered $120^\circ$ structure.

The behaviour of the order parameter ${\cal N}(\alpha,N)$ of the collinear
structure is found to differ between the $N=12$ and $N=24$ systems on the
one hand and the $N=18$ and $N=30$ systems on the other. While
${\cal N}(\alpha,12)$ and  ${\cal N}(\alpha,24)$ increase smoothly as $\alpha$
increases through the classical transition point between the
$120^\circ$ order and the collinear order, ${\cal N}(\alpha,18)$ and
${\cal N}(\alpha,30)$ increase discontinuously at $\alpha_{qu}(18)=0.23$
and $\alpha_{qu}(30)=0.2$ where ${\cal M}(\alpha,18)$ and
${\cal M}(\alpha,30)$ are still finite.
These discontinuities are related to a crossover from a
translationally invariant ground state for $\alpha<\alpha_{qu}$ to a ground
state that has the periodicity of the collinear structure for
$\alpha>\alpha_{qu}$. Thus, for the system sizes $N=18$ and $30$ one recovers
the classical picture of a first order transition between the
$120^\circ$ order and the collinear order. In the cases $N=12$ and $24$ the
ground states remain translationally invariant for all values of $\alpha$.
For these system sizes, the dependence
of ${\cal M}(\alpha,N)$ and ${\cal N}(\alpha,N)$ on $\alpha$ and $N$
agrees with the finite size behaviour of the order parameters of
a system that undergoes a first order
transition between the ordered states characterised by
${\cal M}$ and ${\cal N}$
in the bulk limit.
In summary,
our exact results support the conjecture \cite{Joli2} of a first order
transition between $120^\circ$ order and collinear order. In agreement
with this we see no significant enhancement of the chirality in the transition
region.\\[4cm]

\noindent\hspace*{4.5cm}{\large \bf Acknowledgements}\\
We thank S. Miyashita, W. Apel and M. Wintel
for helpful discussions. This work has been supported the Deutsche
Forschungsgemeinschaft under grant no. Ev 6/3-1.
\begin{appendix}
\renewcommand{\alpheqn}{\setcounter{saveeqn}{\value{equation}}%
\stepcounter{saveeqn}%
\setcounter{equation}{0}%
\renewcommand{\theequation}{\Alph{section}.\mbox{\arabic{saveeqn}
\alph{equation}}}}
\renewcommand{\reseteqn}{\setcounter{equation}{\value{saveeqn}}%
\renewcommand{\theequation}{\Alph{section}.\arabic{equation}}}
\reseteqn
\setcounter{equation}{0}

\newpage
\vspace*{1cm}
\noindent
{\Large \bf Appendix A: Second order spin wave approximation}\\
\addtocounter{section}{1}

\noindent
A straightforward evaluation of the expectation values in
(\ref{2.OrderSWEnergy})
yields the following expressions:
\bea
\frac{\ew{H_2}_0}{N} = -\frac{3}{32S^2} F(b) +  \frac{3}{16S^2} \alpha \,G(b),
\qquad b\equiv\frac{B}{3S}
\eea
where
\bea
F(b)&=&\left(2-2J_1+\frac{1}{N}\sum_{{\bf k}} (b-2\alpha({\bf k}))
\,(A_1({\bf k},B)+A_2({\bf k},B))\right)^2\nonumber\\
&& +B_1^2+4B_1 B_2 -2B_2^2 -(B_1+2B_2)(A_1-A_2),\nonumber\\
G(b)&=&(A_1+A_2-2-C_1-C_2)^2 + (C_1-C_2)(C_1-C_2-A_1+A_2).\nonumber
\eea
and
$$
J_1=\frac{1}{N} \sum_{{\bf k}} \nu({\bf k},B),
$$
with $\nu({\bf k},B)$ as defined in (\ref{SWfrequ}). $A_i, B_i$ and $C_i$
denote the sums
\bea
A_i &=& \frac{1}{N} \sum_{{\bf k}} A_i ({\bf k}),\nonumber\\
B_i &=& \frac{1}{N} \sum_{{\bf k}} \gamma({\bf k}) A_i ({\bf k}),\nonumber\\
C_i &=& \frac{1}{N} \sum_{{\bf k}}\left(1-\frac{\alpha({\bf k})}{\alpha}\right)
A_i ({\bf k}),\qquad i=1,2\nonumber
\eea
where
\bea
A_1({\bf k})=[A_2({\bf k})]^{-1}=
\sqrt{\frac{1+b-2\alpha({\bf k})+2\gamma({\bf k})}
{1+b-2\alpha({\bf k})-\gamma({\bf k})}},\nonumber
\eea
with $\alpha({\bf k}))$ and $\gamma({\bf k}))$ as defined in (\ref{Alphak},b).

Similarly,
\bea
\sum_{\{n_{{\bf k}}\}}
\frac{\bra{ \{n_{{\bf k}}\} }H_{3/2}  \ket{ \{n_{{\bf k}}\} }}
{E(\{0\})-E(\{n_{{\bf k}}\})}=-\frac{3}{16S^2}\frac{1}{N^2} \sum_{{\bf k},
{\bf p}} \frac{I^2({\bf k},{\bf p})}{\nu({\bf k})+\nu({\bf p})+
\nu({\bf k}+{\bf p})},
\eea
where
\bea
I^2({\bf k},{\bf p})&=&A_1({\bf k}) A_1({\bf p}) A_1({\bf k}+{\bf p})
\,[\beta({\bf k})(1-A_2({\bf p}) A_2({\bf k}+{\bf p}))\nonumber\\
&&+\beta({\bf p})(1-A_2({\bf k}) A_2({\bf k}+{\bf p}))\nonumber\\
&&-\beta({\bf k}+{\bf p})(1-A_2({\bf k}) A_2({\bf p}))]^2\nonumber,
\eea
with
$$\beta({\bf k})=-\frac{1}{3}
\left[\sin(k_x)-\sin(\frac{1}{2} k_x-\frac{\sqrt{3}}{2} k_y)
-\sin(\frac{1}{2} k_x+\frac{\sqrt{3}}{2} k_y)\right].
$$
Miyake's result \cite{Miya1} is reproduced by setting $\alpha=0$
in these expressions.

To obtain the second order correction $\chi^y_{2.SW}$ to the helicity we
follow the same procedure as in the determination of ${\cal M}_{2.SW}$.
We replace the Zeeman term in (\ref{SWHamiltonian}) by the term
\bea
-h \sum_{\bigtriangleup} {\hat \chi}_{\bigtriangleup}^y\nonumber,
\eea
which couples the helicity to a fictitious source field $h$ so that
$$
\chi^y = -\left. \frac{1}{N}
\frac{\partial}{\partial h} E_0(\alpha,h)
\right|_{h\to 0},
$$
where $E_0(\alpha,h)$ is the $h$-dependent ground state energy of the
system. The second order term in the expansion of $E_0(\alpha,h)$ with
respect to $S^{-1}$ has the same structure as $E_0(\alpha,B)$,
(\ref{2.OrderSWEnergy}), i.e. it consists of two additive contributions
that contain singular parts proportional to $\sqrt{h}$ which cancel in the
sum. Thus, the limit
$$
\chi_{2.SW}^y = -\frac{1}{N}\,\lim_{h\to 0}
\frac{E_0^{(2)}(\alpha,h)-E_0^{(2)}(\alpha,0)}
{h}
$$
can be performed numerically in exactly the same way as in
(\ref{2.OrdSWMag}).
\reseteqn
\setcounter{equation}{0}

\vspace*{1cm}
\noindent
{\Large \bf Appendix B: \quad \boldmath$ {\cal M}_{1.SW}$ in the limit\,
\boldmath$-\alpha/N\to \infty$}\\
\addtocounter{section}{1}

\noindent
{}From (\ref{Alphak},b) it follows that for
${\bf k}\neq 0,\pm{\bf Q}_1$
\bea
|\alpha({\bf k})|\ge 3\pi^2 \frac{|\alpha|}{N}
\eea
and hence, in the limit $|\alpha|/N\gg 1$,
\bea
|\alpha({\bf k})|\gg 1 > |\gamma({\bf k})|,
\eea
so that $A_{1,2}({\bf k})$ can be expanded in this limit:
\bea
A_{1,2}({\bf k})= 1\mp\frac{1+2\gamma({\bf k})}{2\alpha({\bf k})}\pm
\frac{1-\gamma({\bf k})}{2\alpha({\bf k})}+O(N^2/\alpha^2).
\eea
Using this expansion in (\ref{MagFinSW}) one finds the physically correct
result
\bea
{\cal M}_{1.SW}(\alpha,N)= S + O(N^2/\alpha^2).
\eea
By contrast, the expression
\bea
{\cal M}_{1.SW}(\alpha,N)= S
\left\{1+\frac{1}{4S}\,\left(2-
\frac{1}{N} \sum_{{\bf K}\epsilon BZ \atop {\bf k}\neq 0}
A_1({\bf k})-
\frac{1}{N} \sum_{{\bf K}\epsilon BZ \atop {\bf k}\neq \pm {\bf Q}_1}
A_2({\bf k}))\right)\right\},
\eea
which derives from (\ref{Mag1stSW}) if one neglects only the infinite
contributions of the zero modes, yields the unphysical result
\bea
{\cal M}_{1.SW}(\alpha,N)= S\,\left\{1+\frac{1}{4S} \frac{6}{N} +
 O(N^2/\alpha^2)\right\}.
\eea
Obviously, the finite size correction $-3/(2N)$ which is accounted for in
(\ref{MagFinSW}) is important for systems sizes $N\le 36$ for which exact
numerical results are available.
\reseteqn
\setcounter{equation}{0}

\vspace*{1cm}
\noindent
{\Large \bf Appendix C: Connection between the structure function  and}
{\Large \bf the order parameter}\\
\addtocounter{section}{1}

\noindent
To evaluate $S_N ({\bf Q}_{1})$ in the limit $-\alpha\gg 1$ we start from
the expression
\bea
S_N ({\bf Q}_{1})= 3\left\langle\left(\sum_{i,j \epsilon {\sf A}}^{N/3} -
\sum_{i \epsilon {\sf A}, j \epsilon {\sf B}}^{N/3}\right)
{\bf S}_i  {\bf S}_j\right\rangle_0
\label{AppStruk}
\eea
which follows from (\ref{StrukFkt}) by taking the equivalence of the three
sublattices into account. Since the spins of one sublattice are ferromagnetic
aligned for $-\alpha\gg 1$ we have
\bea
\lim_{\alpha\to -\infty}\ew{{\bf S}_i  {\bf S}_j}=S^2
\quad\mbox{for}\quad i,j \epsilon {\sf A},\, i\ne j.
\label{AppKorr1}
\eea
However,
\bea
\lim_{\alpha\to -\infty}\ew{{\bf S}_i  {\bf S}_j}=S^2 c
\quad\mbox{for}\quad i \epsilon {\sf A},\, j \epsilon {\sf B},
\label{AppKorr2}
\eea
where $c$ can differ from the value $\cos(2\pi/3)=-1/2$, since the spins of
different sublattices may fluctuate against each other. $c$ is fixed by the
condition that the ground state of systems of size $N$ must be a singlet or a
doublet depending on whether $N$ is even or odd:
\bea
\lim_{\alpha\to -\infty} S_{tot} (N)&=&
\lim_{\alpha\to -\infty}\ew{(\sum_i^N {\bf S}_i)^2}\nonumber\\
&=&\frac{N^2}{12}\left(1+2c+\frac{6}{N}\right)=\left\{
\begin{array}{r@{\quad,\quad}l}
0 & N\, \mbox{even},\\
\frac{3}{4}& N\, \mbox{odd}.
\end{array} \right.
\eea
Hence,
\bea
c=\left\{
\begin{array}{l@{\quad,\quad}l}
-\frac{1}{2}-\frac{3}{N} & N\, \mbox{even},\\
-\frac{1}{2}-\frac{3}{N}+\frac{9}{2N^2}& N\, \mbox{odd}.
\end{array} \right.
\label{AppC}
\eea
Using (\ref{AppKorr1}), (\ref{AppKorr2}) and (\ref{AppC}) we find from
(\ref{AppStruk}) the expressions (\ref{MagSize}) of the
main text for $S_N ({\bf Q}_{1})$.
\end{appendix}

\newpage
\hspace{5.5cm}{\Large \bf Figure Captions}
\vspace*{1cm}

{\bf Fig.1}\hfill\parbox[t]{13.5cm}{
Classical ground state configuration in the regime $\alpha<1/8$.
Bold lines: $nn$ bonds; dashed bold line: $nnn$ bond}\vspace*{1cm}

{\bf Fig.2}\hfill\parbox[t]{13.5cm}{
Classical ground state configuration in the regime $1/8<\alpha<1$.
The state is degenerate with respect to the angle $\theta$.}\vspace*{1cm}

{\bf Fig.3}\hfill\parbox[t]{13.5cm}{
The Brillouin zone of the triangular lattice. The arrows indicate the path
which the Bragg vector ${\bf Q}$ of the ground state configuration takes when
$\alpha$ increases:
${\bf Q}={\bf Q}_1$ for $-\infty<\alpha\le 1/8$;
${\bf Q}={\bf Q}_2$ for $1/8<\alpha\le 1$;
for $1<\alpha<\infty$ ${\bf Q}$ moves continuously from ${\bf Q}_2$ towards
the corner of the $\sqrt{3}\times\sqrt{3}$ Brillouin zone.
}\vspace*{1cm}

{\bf Fig.4~a,b}\hfill\parbox[t]{13.5cm}{
First and second order spin wave results: (a) the sublattice magnetisation
${\cal M}$ as a function $\alpha$; (b) the helicity $\chi^y$ as a function of
$\alpha$.}\vspace*{1cm}

{\bf Fig.5}\hfill\parbox[t]{13.5cm}{
Size dependence of the first order spin wave approximation for ${\cal M}$
at $\alpha=0$. $\bigcirc$: ${\cal M}_{1.SW}(\alpha=0,N)$;
full line: leading size correction to ${\cal M}_{1.SW}(0,\infty)$;
dashed line: leading and subleading correction to
${\cal M}_{1.SW}(0,\infty)$}\vspace*{1cm}

{\bf Fig.6~a-f}\hfill\parbox[t]{13.5cm}{
Finite systems studied in this work.
For the $N=18,24$ and $30$ cells which are not $C_{6v}$
symmetric, the triangular lattice has been distorted such that the outer
contours of the cells take the shape of regular hexagons. The degree of
distortion illustrates the deviation from the $C_{6v}$ symmetry.
}\vspace*{1cm}
\newpage
{\bf Fig.7~a-f}\hfill\parbox[t]{13.5cm}{
The square of the order parameters ${\cal M}$ and ${\cal N}$
of the $120^\circ$ structure and of the collinear structure as a function of
$\alpha$ for various system sizes. ${\cal M}_{1.SW}$ is obtained from eq.
(\ref{MagFinSW}). The insets in (a) and (d) show an enlargement of the
transition region.
}\vspace*{1cm}

{\bf Fig.8}\hfill\parbox[t]{13.5cm}{
The difference $\Delta{\cal M}(\alpha,N)={\cal M}(\alpha,N)-
{\cal M}_{1.SW}(\alpha,N)$ and the second order spin wave correction
${\cal M}_{2.SW}(\alpha)$ as functions of $\alpha$.}\vspace*{1cm}

{\bf Fig.9~a-d}\hfill\parbox[t]{13.5cm}{
The square of the helicity $\chi^y$ of the $120^\circ$ structure
as a function of $\alpha$ for various system sizes.
$\chi_{1.SW}$ is obtained from eq. (\ref{HelFinSW}).
}\vspace*{1cm}

{\bf Fig.10}\hfill\parbox[t]{13.5cm}{
The difference $\Delta \chi(\alpha,N)=\chi (\alpha,N)-
\chi_{1.SW}(\alpha,N)$ and the second order spin wave correction
$\chi_{2.SW}(\alpha)$ as functions of $\alpha$.}\vspace*{1cm}

{\bf Fig.11}\hfill\parbox[t]{13.5cm}{
The chiralities $P_{\pm}(\alpha,N)$ as functions of $\alpha$.}\vspace*{1cm}

{\bf Fig.12}\hfill\parbox[t]{13.5cm}{
A plaquette of four spins on which the uniform and staggered chiral operators
are defined.
}
\end{document}